\pdfoutput=1

\documentclass[10pt,journal,compsoc]{IEEEtran}

\ifCLASSOPTIONcompsoc
  \usepackage[nocompress]{cite}
\else
  \usepackage{cite}
\fi
\ifCLASSINFOpdf
\else
\fi
\usepackage{url}
\usepackage[english]{babel}
\usepackage{color}

\usepackage{enumitem} % for more flexible enumerations

\definecolor{mygreen}{rgb}{0,0.6,0}
\definecolor{mygray}{rgb}{0.5,0.5,0.5}
\definecolor{mymauve}{rgb}{0.58,0,0.82}

\usepackage{amssymb}

\usepackage{comment}

\usepackage{epsfig}
\usepackage{array, colortbl}
\usepackage{listings}
\usepackage{epstopdf}
\usepackage{multirow}

\usepackage{float}
\usepackage{balance}
\usepackage{fancybox}
\usepackage{scalefnt}
\usepackage{cleveref}
\usepackage[normalem]{ulem}
\usepackage{bigstrut}
\usepackage{rotating}

\usepackage{leftidx}

\usepackage{tablefootnote}
\usepackage{siunitx}

\usepackage{caption}
\usepackage{graphicx}
\usepackage{mathtools}
 \usepackage{verbatim}

\usepackage{tabularx, booktabs}
\newcolumntype{Y}{>{\centering\arraybackslash}X}

\usepackage{listings}

\usepackage{colortbl}
\usepackage[table*]{xcolor}
\xdefinecolor{gray95}{gray}{0.65}
\xdefinecolor{gray50}{gray}{0.10}
\xdefinecolor{gray25}{gray}{0.8}

\makeatletter
\newcommand{\removelatexerror}{\let\@latex@error\@gobble}
\makeatother

\newcommand{\PQone}{\textbf{What is the relation between anti-patterns and energy efficiency?}}
\newcommand{\PQtwo}{\textbf{What is the relation between anti-pattern types and energy efficiency?}}

\newcommand{\ea}{{et al.}}

\newcommand{\aka}{{\textit{a.k.a.,}}}

\newcommand{\eg}{{\textit{e.g.,}}}
\newcommand{\ie}{{\textit{i.e.,}}}

	\newcommand{\hypobox}[1]{\begin{center}%	
	\noindent\thicklines\setlength{\fboxsep}{8pt}%	
	\cornersize{0.2}\Ovalbox{\begin{minipage}{3.20in}%	
	\textit{#1}\end{minipage}} \end{center}}

\usepackage[ruled,lined,linesnumbered]{algorithm2e}
\SetAlFnt{\small}
\SetAlCapFnt{\small}
\SetAlCapNameFnt{\small}
\SetAlCapHSkip{0pt}
\IncMargin{-\parindent}

\newlist{myenumi}{description}{10}
\setlist[myenumi]{labelindent=\parindent, leftmargin=*, label=(\roman*), align=left}
\setlist[myenumi]{leftmargin=0pt}

\begin{document}

\lstset{
basicstyle=\tiny,
language=Java,
breaklines=true,
commentstyle=\color{mygreen},
keywordstyle=\color{blue},
numberstyle=\tiny\color{mygray},
rulecolor=\color{black},
stringstyle=\color{mymauve},
keepspaces=true,
numbers=left,                    % where to put the line-numbers; possible values are (none, left, right)
 numbersep=5pt,                   % how far the line-numbers are from the code
xleftmargin=0.5cm,frame=tlbr,framesep=2pt,framerule=0pt,
}

%\linenumbers
\title{Anti-patterns and the energy efficiency of Android applications}

\author{Rodrigo~Morales,~\IEEEmembership{Member,~IEEE,}
        Rub\'{e}n~Saborido,~\IEEEmembership{Member,~IEEE,}
				Foutse Khomh,~\IEEEmembership{Member,~IEEE,}
				Francisco Chicano,~\IEEEmembership{Member,~IEEE,}
				and~Giuliano Antoniol,~\IEEEmembership{Senior Member,~IEEE}% <-this % stops a space
\IEEEcompsocitemizethanks{\IEEEcompsocthanksitem R. Morales, R. Saborido, F. Khomh, and G. Antoniol are with Polytechynique Mont\'{e}al, Qu\'{e}bec, Canada. \protect\\
E-mail: \{rodrigo.morales, ruben.saborido-infantes, foutse.khomh\}@polymtl.ca, antoniol@ieee.org.

\IEEEcompsocthanksitem F. Chicano is with University of M{\'a}laga, M{\'a}laga, Spain.
E-mail: chicano@uma.es.}%
\thanks{This work is part of the paper ``EARMO: An Energy-Aware Refactoring Approach for Mobile Apps'' submitted to TSE}}



%\markboth{IEEE TRANSACTIONS ON SOFTWARE ENGINEERING,~Vol.~X, No.~X, September~2016}%
%{Morales \MakeLowercase{\textit{et al.}}: EARMO: An Energy-Aware Refactoring Approach for Mobile Apps}

\markboth{}%
{Morales \MakeLowercase{\textit{et al.}}: Anti-patterns and the energy efficiency of Android applications }




\IEEEtitleabstractindextext{%
\begin{abstract}
The boom in mobile apps has changed the traditional landscape of software development by introducing new challenges due to the limited resources of mobile devices, e.g., memory, CPU, network bandwidth and battery. The energy consumption of mobile apps is nowadays a hot topic and researchers are actively investigating the role of coding practices on energy efficiency. Recent studies suggest that design quality can conflict with energy efficiency. Therefore, it is important to take into account energy efficiency when evolving the design of a mobile app. The research community has proposed approaches to detect and remove anti-patterns (i.e., poor solutions to design and implementation problems) in software systems but, to the best of our knowledge, none of these approaches have included  anti-patterns that are specific to mobile apps and--or considered the energy efficiency of apps. In this paper, we fill this gap in the literature by analyzing the impact of eight type of anti-patterns on a testbed of 59 android apps extracted from F-Droid. First, we (1) analyze the impact of anti-patterns in mobile apps with respect to energy efficiency; then (2) we study the impact of different types of anti-patterns on energy efficiency.
We found that then energy consumption of apps containing anti-patterns and not (refactored apps) is statistically different.  Moreover, we find that the impact of refactoring anti-patterns can be positive (7 type of anti-patterns) or negative (2 type of anti-patterns).  Therefore, developers should consider the impact on energy efficiency of refactoring when applying maintenance activities.
\end{abstract}

}

\maketitle

\IEEEdisplaynontitleabstractindextext

% \fi
\IEEEpeerreviewmaketitle

\IEEEraisesectionheading{\section{Introduction}\label{sec:intro}}


\IEEEPARstart{D}{uring} the last five years, and with the exponential growth of the market of mobile apps~\cite{Anthes:2011:IMA:1995376.1995383}, software engineers have witnessed a radical change in the landscape of software development. From a design point of view, new challenges have been introduced in the development of mobile apps such as the constraints related to internal resources, \eg~CPU, memory, and battery; as well as external resources, \eg~internet access.  Moreover, traditional desired quality attributes, such as functionality and reliability, have been overshadowed by subjective visual attributes, \ie~``flashiness''~\cite{6265078}.


Mobile apps play a central role in our life today. We use them almost anywhere, at any time and for everything; \eg~ to check our emails, to browse the Internet, and even to access critical services such as banking and health monitoring. Hence, their reliability and quality is critical. Similar to traditional desktop applications, mobile apps age as a consequence of changes in their functionality, bug-fixing, and introduction of new features, which sometimes lead to the deterioration of the initial design~\cite{Parnas94-ICSE-SoftwareAging}. This phenomenon known as \textit{software decay}~\cite{eick2001does} is manifested in the form of design flaws or anti-patterns. An example of anti-pattern is the \textit{Lazy class}, which occurs when a class does too little, \ie~has few responsibilities in an app.  A \textit{Lazy class} typically is comprised of methods with low complexity and is the result of speculation in the design and-or implementation stage. Another common anti-pattern is the \textit{Blob}, \aka~\textit{God class}, which is a large and complex class that centralizes most of the responsibilities of an app, while using the rest of the classes merely as data holders. A \textit{Blob class} has low cohesion, and hinders software maintenance, making code hard to reuse and understand. Resource management is critical for mobile apps. Developers should avoid anti-patterns that cause battery drain. An example of such anti-pattern is \emph{Binding resources too early class}~\cite{Gottschalk+2013}. This anti-pattern occurs when a class \emph{switches on} energy-intensive components of a mobile device (\eg~Wi-fi, GPS) when they cannot interact with the user. Another example is the use of \emph{private getters and setters} to access class attributes in a class, instead of accessing directly the attributes. The Android documentation~\cite{AndroidPerformance} strongly recommends to avoid this anti-pattern as virtual method calls are up to seven times more expensive than using direct field access~\cite{AndroidPerformance}.

Previous studies have pointed out the negative impact of anti-patterns on change-proneness~\cite{Khomh2012}, fault-proneness~\cite{taba2013icsm}, and maintenance effort~\cite{CodeSmells_overtime}. In the context of mobile apps, Hecht~\ea~\cite{hecht:hal-01178734} found that anti-patterns are prevalent along the evolution of mobile apps. They also confirmed the observation made by Chatzigeorgiou and Manakos~\cite{chatzigeorgiou2010investigating} that anti-patterns tend to remain in systems through several releases, unless a major change is performed on the system.

Recently, researchers and practitioners have proposed approaches and tools to detect~\cite{moha2010decor,marinescuDetection} and correct~\cite{tsantalis2008jdeodorant} anti-patterns. However, these approaches only focus on object-oriented anti-patterns and do not consider mobile development concerns. One critical concern of mobile apps development is improving energy efficiency, due to the short life-time of mobile device's batteries. Some research studies have shown that behavior-preserving code transformations (\emph{i.e., refactorings}) that are applied to remove anti-patterns can impact the energy efficiency of a program~\cite{SahinPC14,DBLP:conf/seke/ParkHL14,Silva2010}. Hecht~\ea~\cite{Hecht:2016:ESP:2897073.2897100} observed an improvement in the user interface and memory performance of mobile apps when correcting Android anti-patterns, like \emph{private getters and setters}, \emph{HashMap usage} and \emph{member ignoring method}, confirming the need of refactoring approaches that support mobile app developers.

Despite these works on anti-patterns and energy consumption, to the best of our knowledge, there is no study on the impact of anti-patterns in mobile apps. In this paper, we aim to fill this gap by studyng the impact of eight well-known Object-oriented (OO) and Android specific (extracted from Android Performance guidelines~\cite{AndroidPerformance}) anti-patterns on energy efficiency. %from object-oriented systems (OO), and Android Performance guidelines~\cite{AndroidPerformance}.
We use a testbed of 59 open-source android apps extracted from the F-Droid marketplace, an Android app repository. 

The primary contributions of this work can be summarized as follows:
\begin{enumerate}
	\item We perform an empirical study of the impact of anti-patterns on the energy efficiency of mobile apps. We also propose a methodology for a correct measurement of the energy consumption of mobile apps, and compare it with a state-of-the-art approach. Our obtained results provide evidence to support the claim that developer's design choices can improve/decrease the energy efficiency of mobile apps.
	
\end{enumerate}

\textbf{The remainder of this paper is organized as follows:} Section~\ref{sec:background} provides some background information on refactoring, energy measurement of mobile apps. Section~\ref{sec:prelim} presents a case study regarding the impact of anti-patterns on energy efficiency.  In Section~\ref{sec:threats-validity}, we discuss the threats to the validity of our study, while in Section~\ref{sec:relatedWork} we relate our work to the state of the art. Finally, we present our conclusions and highlight directions for future work in Section~\ref{sec:conclusion}.

%\smallsection{Paper organization} The remainder of this paper is organized as follows.
%\Cref{sec:background} provides background details about mobile anti-patterns, and refactoring to improve energy consumption.  \Cref{sec:energyAwareRefactoring} describes our proposed approach while \Cref{sec:caseStudy} describes our experimental methodology.  \Cref{sec:study-results} presents and discusses the results of our experiments.
%\Cref{threats-validity} discloses the threats to the validity of our study.
%\Cref{sec:relatedWork} overviews the related literature.
%Finally, \Cref{sec:conclusion} concludes our work and lays out some directions for future work.
%
\section{Background}
\label{sec:background}

This section presents an overview of the main concepts used in this paper.
%\antoniol{we may fail short of a brief primer on ap at least I will give a few lines  to justify why these ap and not other …}
\subsection{Refactoring}
Refactoring, a software maintenance activity that transforms the structure of a code without altering its behavior~\cite{bourque2014guide}, is widely used by software maintainers to counteract the effects of design decay due to the continuous addition of new functionalities or the introduction of poor design choices, \ie~anti-patterns, in the past~\cite{Parnas94-ICSE-SoftwareAging}.
The process of refactoring requires the identification of places where code should be refactored (\eg~anti-patterns). Developers also have to determine which kind of refactoring operations can be applied to the identified locations. This step is cumbersome, as different anti-patterns can have different impact on the software design. Moreover, some refactoring operations can be conflicting, hence, finding the best combination of refactorings is not a trivial task. More formally, if $k$ is the number of available refactorings, then, the number of possible solutions (NS) is given by $NS=(k!)^k$~\cite{OuniKessentini}, which results in a large space of possible solutions to be explored exhaustively.
Therefore, researchers have reformulated the problem of automated-refactoring as a combinatorial optimization problem and proposed different techniques to solve it. The techniques range from single-objective approaches using local-search metaheuristics, \eg~hill climbing, and simulated annealing~\cite{okeeffeCinneide2006,sengSB06}, to evolutionary techniques like genetic algorithm, and multiobjective approaches: \eg~NSGA-II and MOGA~\cite{simons2010,OuniKessentini,MahouachiKessentiniandCinneide2013,MkaouerKessentiniBechikhCinneide2014}, MOCell, NSGA-II and SPEA2~\cite{Morales2016SanerTesting}.

Recent works \cite{gottschalk2013energy,DBLP:conf/seke/ParkHL14} have provided empirical evidence that software design plays also an important role in the energy consumption of mobile devices; \ie{} high-level design decisions during development and maintenance tasks impact the energy efficiency of mobile apps. More specifically, these research works have studied the effect of applying refactorings to a set of software systems; comparing the energy difference between the original and refactored code.

In this research, we propose an approach for measuring the impact of refactoring mobile apps, on energy efficiency. We target two categories of anti-patterns: (i) anti-patterns that stem from common Object-oriented design pitfalls~\cite{Brown98-AntiPatterns,Fowler99-Refactoring} (\ie{} Blob, Lazy Class, Long-parameter list, Refused Bequest, and Speculative Generality) and (ii) anti-patterns that affect resource usages as discussed by Gottschalk~\cite{gottschalk2013energy} and in the Android documentation~\cite{gottschalk2013energy,AndroidPerformance} (\ie{} Binding Resources too early, HashMap usage, and Private getters and setters). We believe that these anti-patterns occur often and could impact the energy efficiency of mobile apps. In the following subsections, we explain how we measure and include energy consumption in our proposed approach.

\subsection{Energy measurement of mobile apps}\label{energymeasurement}
Energy efficiency, a critical concern for mobile and embedded devices, has been typically targeted from the point of view of hardware and lower-architecture layers by the research community. Energy is defined as the capacity of doing work while power is the rate of doing work or the rate of using energy. In our case, the amount of total energy used by a device within a period of time is the energy consumption. \textit{Energy} (\textit{E}) is measured in \textit{joules} (\textit{J}) while \textit{power} (\textit{P}) is measured in \textit{watts} (\textit{W}). Energy is equal to power times the time period $T$ in seconds. Therefore, $E = P \cdot T$. For instance, if a task uses two watts of power for five seconds it consumes 10 joules of energy.

One of the most used energy hardware profilers is the \textit{Monsoon Power Monitor}\footnote{\url{https://www.msoon.com/LabEquipment/PowerMonitor/}}. It provides a power measurement solution for any single lithium (Li) powered mobile device rated at 4.5 volts (maximum three amps) or lower. It samples the energy consumption of the connected device at a frequency of $5\,kHz$, therefore a measure is taken each 0.2 milliseconds.

%\Foutse{Also describe GreenMiner and explain that we contrasts our measurements with that as well...ideally we should show results using the two measurements setups (one figure/table for each tool and show that they are (or not) consistent!!!} \ruben{Here I introduce some information related to Green Miner. We have to take into account that using green miner 1) we measure the entire system and not the app's related methods, 2) we do not know if scenarios are run successfully and 3) we have to introduce a delay before and after the app starts/finishes to allow Green Miner to have time to work. In addition, they use a different model of phone and an older version of the Android OS.}

In this work energy consumption is measured using a more precise environment. Specifically we use a digital oscilloscope \textit{TiePie Handyscope HS5} which offers the \textit{LibTiePie SDK}, a cross platform library for using \textit{TiePie} engineering USB oscilloscopes through third party software. We use this device because it allows to measure using higher frequencies than the \textit{Monsoon Power Monitor}. The mobile phone is powered by a power supply and, between both, we connect, in series, a \textit{uCurrent}\footnote{\url{http://www.eevblog.com/projects/ucurrent/}} device, which is a precision current adapter for multimeter converting the input current in a proportional output voltage ($V_{out}$). The input current ($I$) is calculated by the \textit{uCurrent} device and, therefore, $I = V_{out}$. Knowing $I$ and the voltage supplied by the power supply ($V_{sup}$), we use the \textit{Ohm's Law} to calculate the power usage ($P$) as $P = V_{sup} \cdot I$. The resolution is set up to 16 bits and the frequency to $125\, kHz$, therefore a measure is taken each eight microseconds. We calculate the energy associated to each sample as $E = P \cdot{T} = P \cdot{(8 \cdot{10^{-6})s}}$. Where $P$ is the power of the smart-phone and $T$ is the period sampling in seconds. The total energy consumption is the sum of the energy associated to each sample.

In our experiments, we used a LG Nexus 4 Android phone equipped with a quad-core CPU, a 4.7-inch screen and running the Android Lollipop operating system (version 5.1.1, Build number LMY47V). We believe that this phone is a good representative of the current generation of Android mobile phones because more than three million have been sold since its release in 2013\footnote{\url{https://goo.gl/6guUpf}}, and the latest version of Android studio includes a virtual device image of it for debugging.

\begin{figure}[htb]
\centering
\includegraphics[scale=0.21]{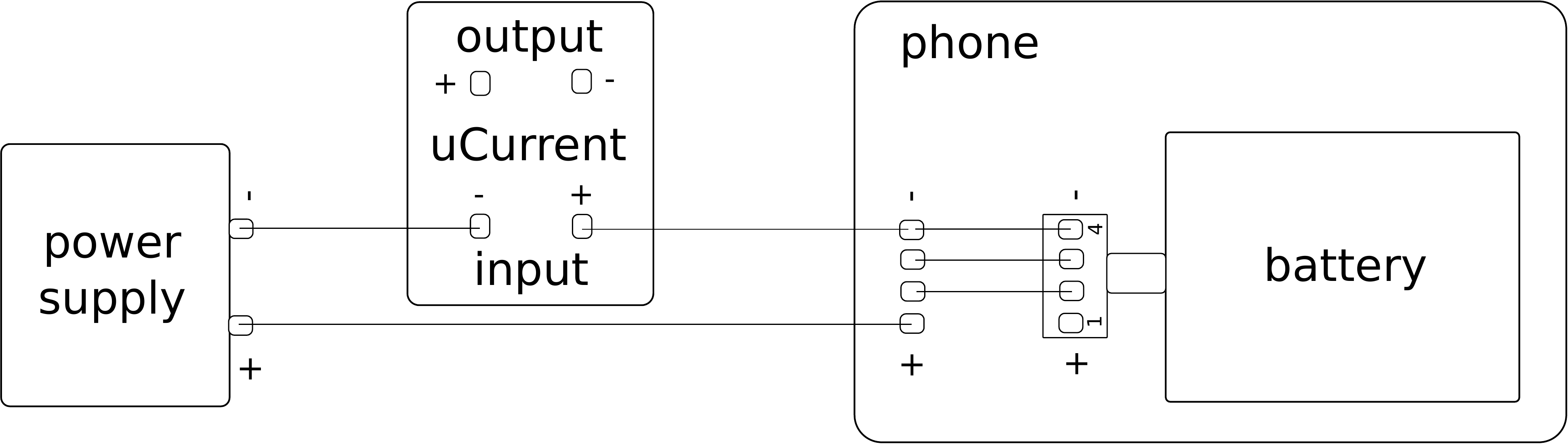}
\caption{Connection between power supply and the Nexus 4 phone. }
\label{fig:circuit}
\end{figure}

We connect the phone to an external power supplier which is connected to the phone's motherboard, thus we avoid any kind of interference with the phone battery in our measurements. The diagram of the connection is shown in \mbox{Fig. \ref{fig:circuit}}.  Note that although we use an external power supplier, the battery has to be connected to the phone to work. Hence, we do not connect the positive pole of the battery with the phone.

To transfer and receive data from the phone to the computer, we use a USB cable, and to avoid interference in our measurements as a result of the USB charging function, we wrote an application to disable it. This application is free and it is available for download in the \textit{Play Store}\footnote{\url{https://goo.gl/wyUcdD}}.

%We leverage the information provided by the aforementioned works as reference for the energy estimation of the refactorings studied in this paper.

%In this research work we propose an energy-aware refactoring framework for assisting software maintainers.
%The refactorings studied in this work fall in the following categories: Object-oriented (OO) techniques like the ones proposed by Fowler \ea~\cite{Fowler99-Refactoring}, and mobile-refactorings like the ones proposed by Gottschalk~\cite{gottschalk2013energy} and in the Android documentation~\cite{AndroidPerformance}.

\section{Case Study}
\label{sec:prelim}
The main goal of this paper is to support developers that aim to improve the design quality of their apps, while controlling for energy consumption. To achieve this goal, the first step is to measure the impact of anti-patterns (\ie{} poor design choices) on energy efficiency. Understanding if anti-patterns affect the energy efficiency of mobile apps is important for practitioners and researchers interested in 1) improving the design of apps through refactoring, and 2) toolsmiths interested in develop tools to automatically improve the design of an app, while performing regular coding tasks.  Specifically, if anti-patterns do not significantly impact energy consumption, then it is not necessary to control for energy efficiency during a refactoring process. Whereas, if anti-patterns significantly affect energy consumption, developers and practitioners should be equipped with refactoring approaches that control for energy efficiency during the refactoring process, in order to prevent a deterioration of the energy efficiency of apps. \\%if there is a significant impact on energy efficiency of classes with anti-patterns, we can support developers and practitioners by providing an approach that suggests refactorings to improve both dimensions, design quality and energy efficiency.\\

We formulate the research questions of this case study as follows:
%\vspace{-15pt}
%\begin{itemize}
\begin{myenumi}%[{\bf (RQ1)}]
	\item[\bf{(RQ1)}] \PQone \\
	The rationale behind this question is to determine if the energy efficiency of mobile apps with anti-patterns differs from the energy efficiency of apps without anti-patterns. We test the following null hypothesis:
	\emph{$H_{0_1}$: there is no difference between the energy efficiency of apps containing anti-patterns and apps without anti-patterns}.
	\item[\bf{(RQ2)}] \PQtwo \\
	In this research question, we analyze whether certain types of anti-patterns lead to more energy consumption than others. We test the following null hypothesis: \emph{$H_{0_2}$: there is no difference between the energy efficiency of apps containing different types of anti-patterns.} %does not differs from others
	
	%We analyze whether certain kind of a leads to more energy consumption than others by testing the null hypothesis: $H_{0_2}$: the energy consumption of a refactored app is not affected by the kind of refactoring applied
	%
	%We analyze whether certain anti-patterns imply more energy consumption than others by testing the null hypothesis: $H_{0_2}$: the energy consumption of apps participating in certain anti-patterns is not higher than others
	%
	
%\end{itemize}
\end{myenumi}

\subsection{Design of the Study}
\label{subsec:prelimStudy}
%talk about the criteria of the designed apps

As mentioned earlier, we consider two categories of anti-patterns: (i) \emph{Object-oriented (OO)} anti-patterns~\cite{Brown98-AntiPatterns,Fowler99-Refactoring}, and (ii) \emph{Android anti-patterns (AA)} defined by~\cite{gottschalk2013energy,AndroidPerformance}. \mbox{Table \ref{table:defects}} presents the details of these anti-patterns. We select these anti-patterns because they have been found in mobile apps~\cite{hecht:hal-01178734,Hecht:2016:ESP:2897073.2897100}, and they are well defined in the literature with recommended steps to remove them~\cite{Fowler99-Refactoring,Brown98-AntiPatterns,gottschalk2013energy,AndroidPerformance}.

Some of the refactorings applied to remove the aforementioned anti-patterns have been previously evaluated in terms of energy consumption using software estimation approaches. For example, \emph{Binding resources too early} was evaluated by Gottschalk~\cite{gottschalk2013energy} and Park~\ea~\cite{DBLP:conf/seke/ParkHL14} evaluated the refactorings proposed by Fowler. For Android anti-patterns like \emph{HashMap usage}, and \emph{private getters and setters}, there is no energy-consumption evaluation that we are aware of, however, they have been reported to decrease memory performance in previous works~\cite{Hecht:2016:ESP:2897073.2897100}. We believe that these anti-patterns occur often in mobile apps and could impact their energy efficiency.

\begin{table}[!t]
\centering

\renewcommand{\arraystretch}{1.0}
\renewcommand{\tabcolsep}{.5mm}
\caption{List of studied Anti-patterns.}
\label{table:defects}
\scriptsize
\begin{tabular}{|p{0.2\columnwidth}|p{0.43\columnwidth}|p{0.34\columnwidth}|}
\hline
\cellcolor{gray95}Name &\cellcolor{gray95} Description &\cellcolor{gray95} Refactoring(s) strategy\tabularnewline
\hline
\multicolumn{3}{|c|}{\cellcolor{gray25}Object-oriented anti-patterns}\tabularnewline
\hline
Blob (BL)~\cite{Brown98-AntiPatterns} & A large class that absorbs most of the functionality of the system with very low cohesion between its constituents. & \textit{Move method (MM)}.  Move the methods that does not seem to fit in the Blob class abstraction to more appropriate classes~\cite{sengSB06}.\tabularnewline
\hline
Lazy Class (LC)~\cite{Fowler99-Refactoring} & Small classes with low complexity that do not justify their existence
in the system. & \textit{Inline class (IC)}. Move the attributes and methods of the LC to another class in the system.\tabularnewline
\hline
Long-parameter list (LP)~\cite{Fowler99-Refactoring} & A class with one or more methods having a long list of parameters. & \textit{Introduce parameter object (IPO)}. Extract a new class with the long list of parameters and replace the method signature. \tabularnewline
\hline
Refused Bequest (RB)~\cite{Fowler99-Refactoring} & A subclass uses only a very limited functionality of the parent class. &  \textit{Replace inheritance with delegation (RIWD)}. Remove the inheritance from the RB class and replace it with delegation through using an object instance of the parent class.\tabularnewline
\hline
%Spaghetti Code (SC)~\cite{Brown98-AntiPatterns} & A class without structure that declares long methods without parameters
%& \textit{Extract Super Class, Replace method with method object}. Extract long methods to new classes and extract a super class with the attributes and methods shared by the SC class and the new extracted classes\tabularnewline
%\hline
Speculative Generality (SG)~\cite{Fowler99-Refactoring} & There is an abstract class created to anticipate further features, but it is only extended by one class adding extra complexity to the design. & \textit{Collapse hierarchy (CH)}. Move the attributes and methods of the child class to the parent and remove the \textit{abstract} modifier. \tabularnewline
\hline
\multicolumn{3}{|c|}{\cellcolor{gray25} Mobile anti-patterns}\tabularnewline
\hline
Binding Resources too early (BE)~\cite{gottschalk2013energy} & Refers to the initialization of high-energy-consumption components of the device, \eg~GPS, Wi-Fi before they can be used. & \textit{Move resource request to visible method (MRM)}. Move the method calls that initialize the devices to a suitable Android event. For example, move method call for \texttt{requestlocationUpda\-tes}, which starts GPS device, after the device is visible to the app/user (\texttt{OnResume} method).\tabularnewline
\hline
HashMap usage (HMU)~\cite{Hecht:2016:ESP:2897073.2897100} & From API 19, Android platform provides \emph{ArrayMap}\tablefootnote{\url{https://developer.android.com/reference/android/support/v4/util/ArrayMap.html}} which is an enhanced version of the standard \emph{Java HashMap} data structure in terms of memory usage. According to Android documentation, it can effectively reduce the growth of the size of these arrays when used in maps holding up to hundreds of items. &  \textit{Replace HashMap with ArrayMap (RHA)}. Import ArrayMap and replace HashMap declarations with ArrayMap data structure. \tabularnewline
\hline
Private getters and setters (PGS)~\cite{Hecht:2016:ESP:2897073.2897100,AndroidPerformance} & Refers to the use of private getters and setters to access a field inside a class decreasing the performance of the app because of simple inlining of Android virtual machine~\tablefootnote{\url{https://source.android.com/devices/tech/dalvik/}} that translates this call to a virtual method called,  which is up to seven times slower than direct field access. & \textit{Inline private getters and setters (IGS)}. Inline the private methods and replace the method calls with direct field access.\tabularnewline
\hline
%Releasing resources too late (RL)~\cite{gottschalk2013energy} & Refers to neglecting to turn-off high-energy consumption components of the device after they are no longer used & Move the methods that release the use of resources to a suitable Android event \eg~move the method removeUpdates that disables GPS, from the closing event of the app (OnDestroy) to the event when the app lost the focus (OnPause) \tabularnewline \hline

\end{tabular}
%\vspace{-10pt}
\end{table}

To study the impact of the anti-patterns, we randomly downloaded 59 android apps from F-droid, an open-source Android app repository\footnote{\url{https://f-droid.org/}}. These apps come from five different categories (Games, Science and Education, Sports and health, Navigation, and Multimedia). To select the apps used in our study, we set the following criteria: more than one class, with at least one instance of any of the anti-patterns studied. Because we physically measure the energy consumption of the apps on a real phone, we validate that the candidate app compiles and run in the phone employed in this study.  After discarding the apps that do not respect the selection criteria, we end-up with a dataset of 20 apps.  \mbox{Table \ref{table:apps} shows the selected apps.}

\begin{table*}[!t]
\renewcommand{\arraystretch}{1.0}
\renewcommand{\tabcolsep}{.5mm}
%\begin{small}
\centering
\caption{Apps used to conduct the case study.}
\label{table:apps}
%\scriptsize
%\begin{tabular}{p{0.8in}p{0.3in}p{0.3in}p{0.6in}p{0.9in}}
\begin{tabular}{lllll}
\hline
App              & Version & LOC & Category             & Description
 \\
 \hline
blackjacktrainer & 0.1     & 3783 	& Games                & Learning BlackJack                       \\
calculator       & 5.1.1   & 13985 	& Science \& Education & Make calculations                        \\
gltron           & 1.1.2   & 12074 	& Games                & 3D lightbike racing game                 \\
kindmind         & 1.0.0   & 6555		& Sports \& Health     & Be aware of sad feelings and unmet needs \\
matrixcalc       & 1.5     & 2416		& Science \& Education & Matrix calculator                        \\
monsterhunter    & 1.0.4   & 27368	& Games                & Reference for Monster Hunter 3 game      \\
mylocation       & 1.2.1   & 1146		& Navigation           & Share your location                      \\
oddscalculator   & 1.2     & 2226		& Games                & Bulgarian card game odds calculator      \\
prism            & 1.2     & 4277		& Science \& Education & Demonstrates the basics of ray diagrams  \\
quicksnap        & 1.0.1   & 18487	& Multimedia           & Basic camera app                         \\
SASAbus          & 0.2.3   & 9349		& Navigation           & Bus schedule for South Tyrol             \\
scrabble         & 1.2     & 3165		& Games                & Scrabble in french									     \\
soundmanager     & 2.1.0   & 5307		& Multimedia           & Volume level scheduler                   \\
speedometer      & 1       & 139		& Navigation           & Simple Speedometer                       \\
stk              & 0.3     & 4493		& Games                & A 3D open-source arcade racer            \\
sudowars         & 1.1     & 22837	& Games                & Multiplayer sudoku                       \\
swjournal        & 1.5     & 5955		& Sports \& Health     & Track your workouts                      \\
tapsoffire       & 1.0.5   & 19920	& Games                & Guitar game                              \\
vitoshadm        & 1.1     & 567		& Games                & Helps you to make decisions              \\
words            & 1.6     & 7125		& Science \& Education & Helps to study vocabulary for IELTS exam            \\
\hline
\end{tabular}
%\end{small}
\end{table*}

\subsection{Data Extraction}
\label{subsec:data-extraction}
\begin{figure*}[htp]
%\centering
\centerline{
\includegraphics[scale=0.55]{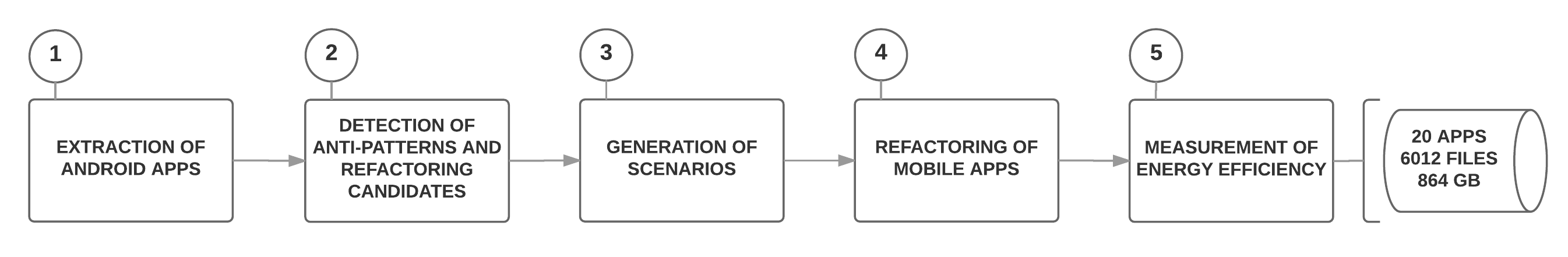} %for resizing the figure
}
\caption{Data extraction process.}
\label{fig:dataExtractionDiagram}
\end{figure*}

The data extraction process is comprised of the following steps, which are summarized in \mbox{Fig. \ref{fig:dataExtractionDiagram}}.

%\begin{description}

\begin{enumerate}

%\item[Step 1: Extract android apps]
\item \textbf{Extraction of android apps.}
We wrote a script to download the apps from \emph{F-droid} repository. This script provides us with the name of the app, the link to the source code, Android API version, and the number of Java files. We use the API version to discriminate apps that are not compatible with our phone, and the number of java files to filter apps with only one class. After filtering the apps, we import the source code in Eclipse (for the older versions) or Android Studio and ensure that they can be compiled and executed.

\item \textbf{Detection of  anti-patterns and refactoring candidates}.
The detection and generation of refactoring candidates is performed using our previous automated approach \emph{ReCon}~\cite{Morales2016Recon}. We use ReCon's current implementation of object-oriented anti-patterns and add two new OO anti-patterns (\textit{Blob} and \textit{Refused bequest}); we also add four Android anti-patterns based on the guidelines defined by Gottschalk~\cite{gottschalk2013energy}, and the \emph{Android documentation}~\cite{AndroidPerformance}. ReCon supports two modes, root-canal- and floss-refactoring. We use the root-canal mode as we are interested in improving the complete design of the studied apps.

\item \textbf{Generation of scenarios}.
For each app we define a typical usage scenario interacting with each application under study using the Android application \emph{HiroMacro\footnote{https://play.google.com/store/apps/details?id=com.prohiro.macro}}. This software allows us to generate scripts containing touch and move events, imitating a real user interacting with the app on the phone, to be executed several times without introducing variations in execution time due to user fatigue, or skillfulness. To automatize the measurement of the studied apps we convert the defined scenarios (\emph{HiroMacro} scripts) to \textit{Monkeyrunner} format. Thus, the collected actions can be played automatically from a script using the \textit{Monkeyrunner}\footnote{http://developer.android.com/tools/help/monkeyrunner\_concepts.html} Android tool.

\item \textbf{Refactoring of mobile apps}.
%From the list of refactoring candidates, we apply one refactoring per app, in three different apps.

We use \emph{Android Studio} and Eclipse refactoring-tool-support for applying the refactorings suggested by \textit{ReCon}. For the cases where there is no tool support, we applied the refactorings manually into the source code. Currently, there is no tool support for refactoring \emph{Binding resources too early} and \emph{Hashmap usage}.  To be sure that the refactored code is executed in the scenario, we set breakpoints and validate that the debugger stops on it.  We also check that the refactored methods appeared in the execution trace. To activate the generation of execution trace file, we use the methods provided in \emph{Android Debug Class}\footnote{https://developer.android.com/reference/android/os/Debug.html}, for both original and refactored versions. The trace file contains information about all the methods executed with respect to time, that we use in the next step.

\item \textbf{Measurement of energy consumption}. As we mention in Section \ref{sec:background}, we measure energy consumption of mobile apps using a precise digital oscilloscope \textit{TiePie Handyscope HS5} which allows us to measure using high frequencies.

%Specifically we obtain an energy sample each eight microseconds (it supposes 125,000 samples per second). We use a LG Nexus 4 Android phone, equipped with Android Lollipop operating system (version 5.1.1, Build number LMY47V), which is connected to an external power supplier.

In our experiments each app is run 30 times to get median results and, for each run, the app is uninstalled after its usage and the cache is cleaned. A description of the followed steps is given in Algorithm~\ref{alg:script}, which has been implemented as a \textit{python} script. As it is described, all apps are executed before a new run is started. Thus, we aim to avoid that cache memory on the phone stores information related to the app run that can cause to run faster after some executions. In addition, before the experiments, the screen brightness is set to the minimum value and the phone is set to keep the screen on. In order to avoid any kind of interferences during the measurements, only the essential Android services are run on the phone (for example, we deactivate WiFi if the app does not require it to be correctly executed, etc.).

When the oscilloscope is started it begins to store in memory energy measurements which are written to a \textit{Comma Separated Values} (CSV) file when the scenario associated to the app finishes. In addition to energy, the generated file contains a timestamp for each sample. Once Algorithm \ref{alg:script} finishes, we have two files for each app and run: the energy trace and the execution trace. Using the existing timestamp in energy traces and the starting and ending time of methods calls in execution traces, energy consumption is calculated for each method called and this information is saved in a new CSV file for each app and run. From these files, we filtered out  method names that does not belong to the \emph{namespace} of the app. For example, for the Android \emph{calculator} app, the main activity is located in the package \texttt{com.android2.calculator3}, and we only consider the methods included in this package as they correspond to the source code that we analyze to generate refactoring opportunities. This is done to reduce the noise of OS native processes running in the background, and third-party services. Finally, the median and average energy consumption of each app over the 30 runs is calculated.

%To reduce the noise of OS native process running in the background, and third-party services, we filter out the methods that does not belong to the app's package. For example, for the Android \emph{calculator}, the main activity is located in the package \texttt{com.android2.calculator3}, and we only consider the methods included in this package as they correspond to the source code that we analyze to generate refactoring opportunities.

%Once the data have been collected, we processed them and save them as \textit{Comma Separated Values} (CSV) file per each independent run.  Each CSV file contains the mean and median energy consumption of each app.  Each row includes the energy consumption organized per method, the starting and ending time.
\end{enumerate}

\begin{algorithm}[!htbp]
%\begin{scriptsize}
 	\ForAll{$runs$}
 	{
 		\ForAll{$apps$}
 		{
  			Install app (using \textit{adb}). \\
	        Start oscilloscope to measure energy. \\
  			Run app (using \textit{adb}). \\
			Play scenario (using \textit{Monkeyrunner}). \\
			Stop oscilloscope. \\
			Download the execution trace file (using \textit{adb}). \\
			Stop app (using \textit{adb}).\\
			Clean app files (using \textit{adb}). \\		
			Uninstall app (using \textit{adb}). \\
			
		}
	}
%\end{scriptsize}
 \caption{Steps to collect energy consumption.}
 \label{alg:script}
\end{algorithm}

\subsection{Data Analysis and Discussion}
%\begin{enumerate}
%\item[\textbf{(PQ1):}] \PQone\\
In the following we describe the dependent and independent variables of this case study, and the statistical procedures used to address each research question. For all statistical tests, we assume a significance level of 5\%.
In total we collected 864 GB of data from which 391 GB correspond to energy traces, 329 GB to execution traces. The amount of data generated from computing the energy consumption of methods calls using these traces is 144 GB.
%\\

\textbf{(RQ1):} \PQone\\
For \textbf{RQ1}, the \emph{dependent variable} is the energy consumption for each app version (original, refactored). The \emph{independent variable} is the existence of any of the anti-patterns studied, and it is true for the original design of the apps we studied, and false otherwise.
We statistically compare the energy consumption between the original and refactored design using a non-parametric test, Mann-Whitney U test. Because we do not know beforehand if the energy consumption will be higher in one direction or in the other, we perform a two-tailed test. For estimating the magnitude of the differences of means between original and refactored designs, we use the non-parametric effect size measure Cliff's $d$, which indicates the magnitude of the effect size of the treatment on the dependent variable. The effect size is small for 0.147 $\leq d <$ 0.33, medium for 0.33 $\leq d <$ 0.474, and large for $d \geq$ 0.474.

\textbf{(RQ2):} \PQtwo\\
For \textbf{RQ2}, we follow the same methodology as \textbf{RQ1}. For each type of anti-pattern, we have three different apps containing an instance of the anti-pattern. We refactor these apps to obtain versions without the anti-pattern. We measure the energy consumption of the original and refactored versions of the apps 30 times to obtain the values of the \emph{dependent variable}. The \emph{independent variable} is the existence of the type of anti-pattern. %of both the original versions of the apps (containing the anti-pattern) and the refactored versions the apps. Next, we follow the same methodology as PQ1 to answer PQ2. %We compare the difference between%from the apps, and measuredobtaining three
%\rodrigo{To observe if there is any difference in the energy consumption of apps containing different types of anti-patterns, we measure the energy consumption of three different occurrences of each anti-pattern type, in three different apps, and their refactored version.  We report the results following the same methodology as PQ1.}
%To observe if there is any difference in the energy consumption of apps containing different types of anti-patters, we measure three different instances of each anti-pattern \Foutse{which instances exactly? this is not clear...we should explain this more...} from our testbed, and report the results following the same methodology as PQ1.

\subsection{Results of the case Study}

\begin{figure*}[htp]
%\centering
\centerline{
\includegraphics[scale=0.5]{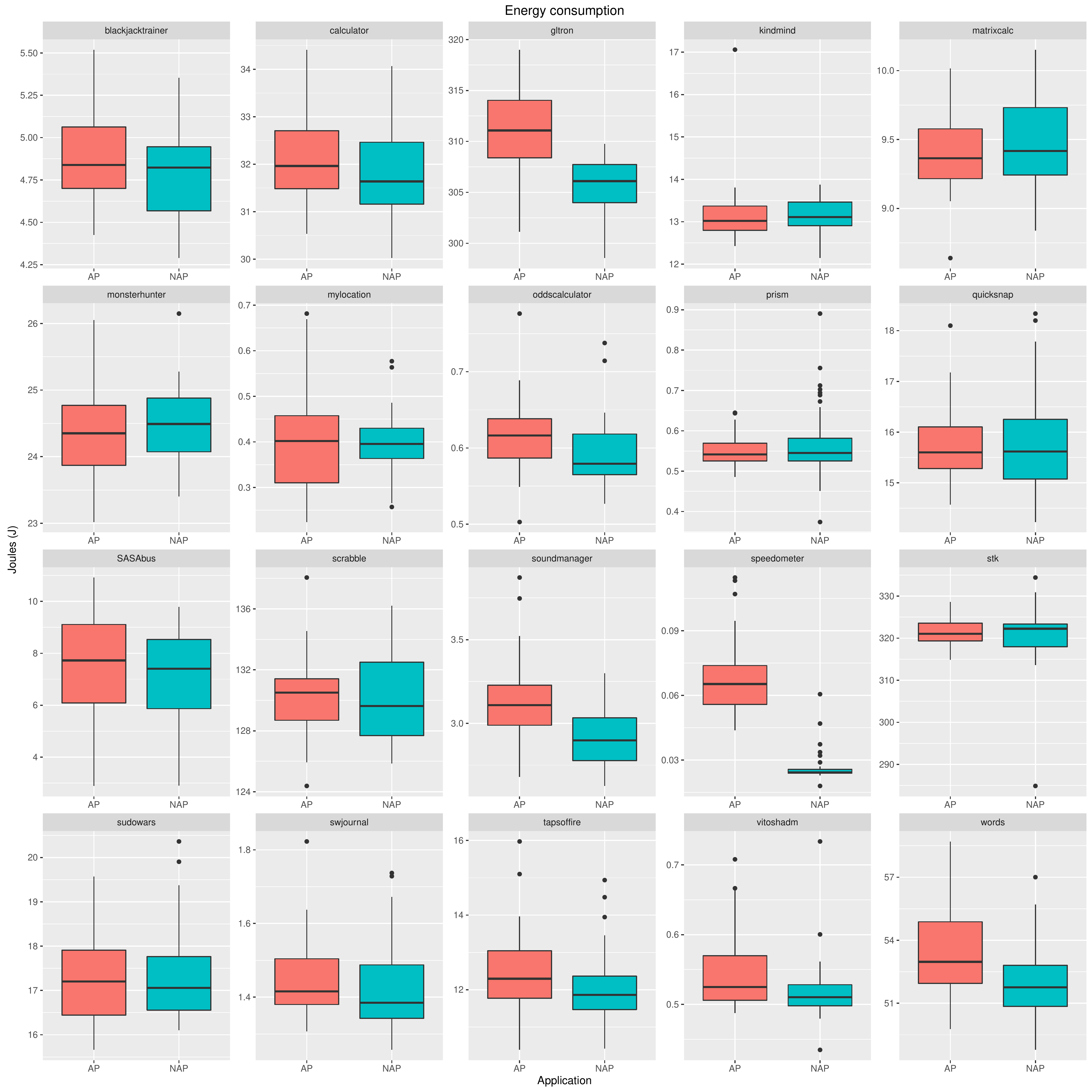} %for resizing the figure
}
\caption{Energy consumption of apps with ($AP$) and without anti-patterns ($NAP$)}
\label{fig:EnergyByApp}
\end{figure*}

In \mbox{Figure \ref{fig:EnergyByApp}} we present the distribution of energy consumption for apps participating in anti-patterns $AP$ and their refactored version $NAP$. We observe that removing anti-patterns in an app can sometimes have a negative impact on the energy efficiency of the app (see the results of \textit{kindmind}, \textit{matrixcalc}, \textit{monsterhunter}). In the 18 remaining apps, the energy consumption is lower in apps without anti-patterns compare to apps with anti-patterns. This result suggests that developers should be careful when removing anti-patterns to improve the design quality of their apps as the operation can have an undesirable effect on energy efficiency (\eg{} it's the case for \textit{kindmind}, \textit{matrixcalc}, \textit{monsterhunter}). This finding is consistent with a previous finding by Sahin~\ea~\cite{SahinPC14}, that refactoring do not always lead to an improvement of the energy efficiency.

In the studied apps we corrected 24 anti-patterns in total. In seven cases (\ie{} 30\%) we obtained a statistically significant difference between the energy consumption of the original and refactored versions of the apps, with Cliff's $\delta$ effect sizes ranging from small to large. Specifically, we obtained three apps with large effect size: \textit{speedometer}, \textit{gltron}, and \textit{soundmanager} (2 type of anti-patterns); two with medium effect size: \textit{oddscalculator}, \textit{words}; and one with small effect size, \textit{vitoshadm}. Therefore we reject $H_{0_1}$ for these seven apps.\\
\hypobox{Overall, our results suggest that different types of anti-patterns may impact the energy efficiency of apps differently. Our next research question (\ie{} \textbf{RQ2}) investigates this hypothesis in more details.} % in the next research question. %, we investigate this hypoth

\begin{figure*}[htp]
%\centering
\centerline{
\includegraphics[scale=0.5]{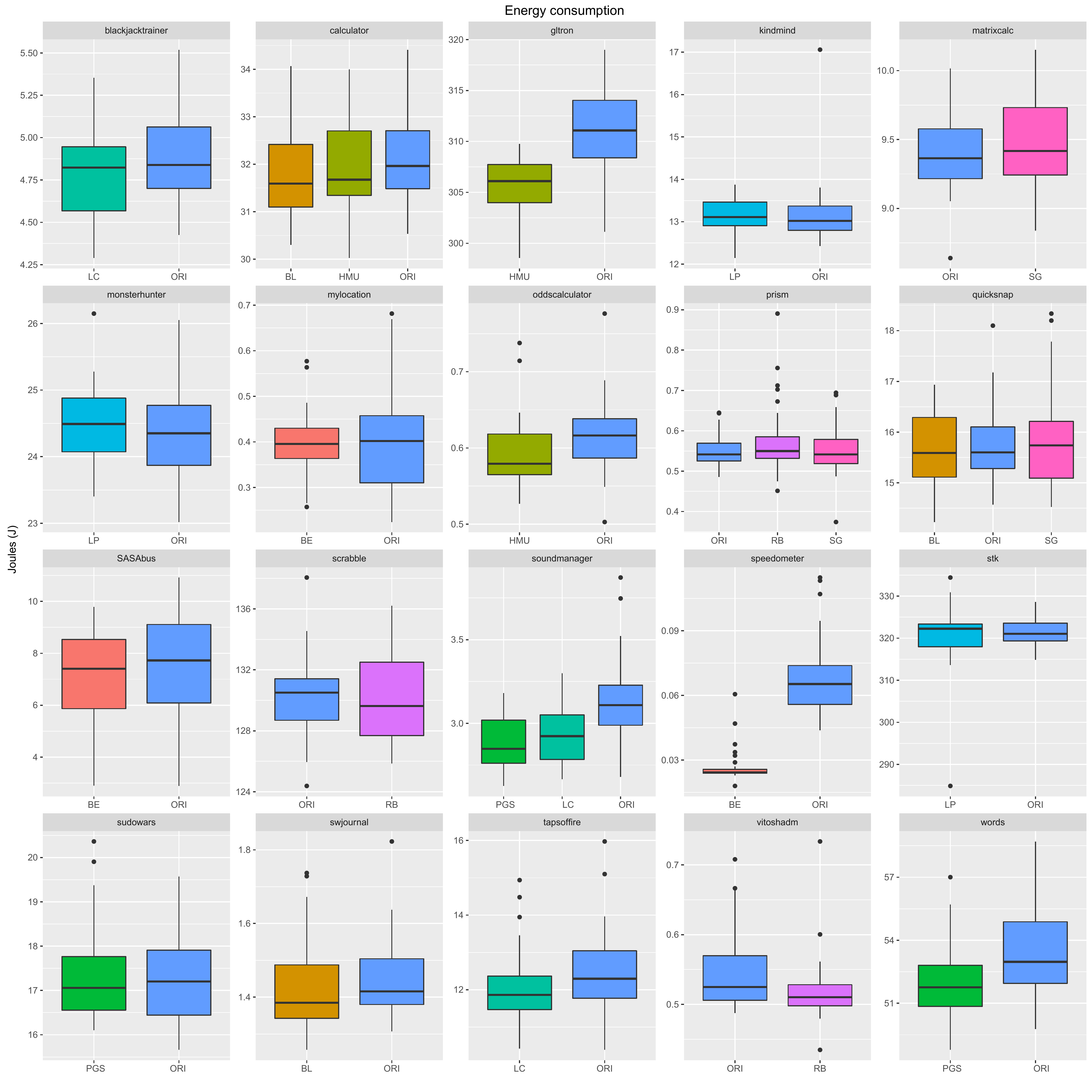} %for resizing the figure
}
\caption{Energy consumption of apps after removing different types of anti-patterns}
\label{fig:EnergybyAntipattern-app}
\end{figure*}

To answer \textbf{RQ2}, on the impact of different types of anti-patterns on energy efficiency, we present in \mbox{Figure \ref{fig:EnergybyAntipattern-app}} the distribution of the energy consumption for each anti-pattern studied, and in \mbox{Table \ref{table:antipatternDetailsMW}} the results of the Mann-Whitney U test and Cliff's $\delta$ effect sizes. %In \Cref{fig:EnergybyAntipattern-app}, ORI correspond to \Foutse{...please explain the symbols used in the figure...}\\ %for each instance studied.

\textbf{Regarding object-oriented (OO) anti-patterns.} In the first plot (position 1, 1 corresponding to \emph{blackJacktrainer}) of \mbox{Figure \ref{fig:EnergybyAntipattern-app}}, we have the original version (ORI), and a refactored version when we remove a \textit{Lazy class} instance (LC). We observe that the median is slightly higher for the original code in comparison with the refactored version. This trend holds for \emph{tapsoffire} (4, 3) and \emph{soundmanager} (3, 3) respectively, with the former one having statistically significance and large magnitude (ES). In the case of \textit{Refused Bequest} (RB), two out of three apps show that removing the anti-pattern saves energy, and the difference is statistically significant for \emph{vitoshadm}. A similar trend is observed for the \emph{Blob}; two out of three apps report a decrease in energy consumption after removing the \emph{Blob}, though the differences are not statistically significant.

\begin{table}[!t]
\renewcommand{\arraystretch}{1.0}
\renewcommand{\tabcolsep}{.5mm}
\caption{Statistical tests for the difference in energy consumption of apps containing different types of anti-patterns. Mann\textemdash Whitney U Test and  Cliff$'$S $\delta$ Effect Size (ES).}
\label{table:antipatternDetailsMW}
\centering
\begin{tabular}{lllll}
\hline
Application&Type&$p-$value&ES&ES Magnitude\tabularnewline
\hline
mylocation&BE&0.57&0.03&small\tabularnewline
SASAbus&BE&0.23&-0.13&small\tabularnewline
speedometer&BE&\textbf{\textless0.05}&\textbf{-0.97}&\textbf{large}\tabularnewline
calculator&BL&0.58&-0.13&small\tabularnewline
quicksnap&BL&0.95&-0.03&small\tabularnewline
swjournal&BL&0.23&-0.23&small\tabularnewline
calculator&HMU&0.43&-0.1&small\tabularnewline
gltron&HMU&\textbf{\textless0.05}&\textbf{-0.7}&\textbf{large}\tabularnewline
oddscalculator&HMU&\textbf{\textless0.05}&\textbf{-0.34}&\textbf{medium}\tabularnewline
soundmanager&IGS&\textbf{\textless0.05}&\textbf{-0.63}&\textbf{large}\tabularnewline
sudowars&IGS&0.64&0.04&small\tabularnewline
words&IGS&\textbf{\textless0.05}&\textbf{-0.44}&\textbf{medium}\tabularnewline
blackjacktrainer&LC&0.4&-0.12&small\tabularnewline
soundmanager&LC&\textbf{\textless0.05}&\textbf{-0.53}&\textbf{large}\tabularnewline
tapsoffire&LC&0.36&-0.22&small\tabularnewline
kindmind&LP&0.3&0.16&small\tabularnewline
monsterhunter&LP&0.34&0.1&small\tabularnewline
stk&LP&0.5&0.02&small\tabularnewline
prism&RB&0.09&0.17&small\tabularnewline
scrabble&RB&0.98&-0.04&small\tabularnewline
vitoshadm&RB&\textbf{\textless0.05}&\textbf{-0.29}&\textbf{small}\tabularnewline
matrixcalc&SG&0.49&0.09&small\tabularnewline
prism&SG&0.72&0.03&small\tabularnewline
quicksnap&SG&0.49&0.04&small\tabularnewline

\hline
\end{tabular}
\end{table}

Concerning \emph{Long Parameter list} (LP), and \emph{Speculative Generality} (SG), both report a negative impact in energy efficiency after refactoring. While for LP, all the apps point toward more energy consumption, in the case of SG, the energy consumption is increased in two out of three apps after refactoring. We explain the result obtained for LP by the fact that the creation of a new object (\ie{} the parameter object that contains the long list of parameters) adds to some extent more memory usage. For SG we do not have a plausible explanation for this trend. For both anti-patterns, the obtained differences in energy consumption is not statistically significant, hence we cannot conclude that these two anti-patterns always increase or decrease energy consumption. %Given that for both types the results are not statistically significant, we could not draw any conclusion.

\textbf{Regarding Android anti-patterns}. For \emph{HashMap usage} (HMU) and \emph{Private getters and setters} (PGS), we obtained statistically significant results for two apps. For \emph{Binding Resources too early} (BE), the result is statistically significant for one app. In all cases, apps that contained these anti-patterns consumed more energy than their refactored versions that did not contained the anti-patterns. This finding is consistent with the recommendation of previous works (\ie{} \cite{Gottschalk+2013,AndroidPerformance}) that advise to remove HMU, PGS, and BE from Android apps, because of their negative effects on energy efficiency. Note that the amount of energy saved is influenced by the context in which the application runs. For example, \textit{SASAbus} is a bus schedule application, and every time we launch the app it downloads the latest bus schedule, consuming a considerable amount of data and energy. As a result, the gain in energy for relocating the call method that starts the GPS device is negligible in comparison to the overall scenario. \emph{Mylocation} is a simpler app, that only provides the coordinated position of mobile user. This app optimizes the use of the GPS device by disabling several parameters, like altitude and speed. It also sets the precision to \emph{coarse} (approximate location\footnote{\url{https://developer.android.com/guide/topics/location/strategies.html}}), and the power requirements to \emph{low}. For this app, we observe a consistent improvement when the anti-pattern is removed, but in a small amount. On the other hand, we have \textit{speedometer}, which is a simple app as well, that measures user's speed, but using \emph{high precision mode}. \emph{High precision mode} uses GPS and internet data at the same time to estimate location with high accuracy. In \textit{speedometer}, we observe a high reduction in energy consumption when the anti-pattern is corrected, in comparison with the previous two apps. %}

%\antoniol{I see a big risk here that the android ap are driving the game and thus it make no sense to consider the other in a way to be the finding is paying against us. I mean in first approximation if you fix energy ap how much to you gain and if you fix the other ,,,, …. feel kind of uneasy. …hmm …. if we consider the Gottschalk and Jelschen chapter is this  a surprise ? }

%\rodrigo{Well the results are more statistically significant for Android smells, though OO smells show the same trend.  Also, consider that Gottshalk didn't measure IGS or Hashmap usage, we took this from the work of Hecht, but there is no paper that evaluate them in terms of energy.}
\hypobox{In summary, removing \textbf{Lazy class}, \textbf{Refused Bequest}, \textbf{Blob}, \textbf{Binding Resources too early}, \textbf{Private getters and setters}, and \textbf{Hashmap usage} anti-patterns can improve the energy efficiency of an Android app (with the removal of the last three anti-patterns providing the biggest savings), while removing \textbf{Long Parameter list}, and \textbf{Speculative Generality} anti-patterns can deteriorate the energy efficiency of the app.} The impact of different types of anti-patterns on the energy consumption of mobile apps is not the same. Hence, we reject $H_{02}$. % (SG). However,  with the
%\hypobox{We reject the null hypothesis $H_{02}$ as the impact on energy efficiency of apps with anti-patterns is different according to the type.  We conclude that Apps participating in Android anti-patterns, \ie~Binding Resources too early, Private getters and setters and Hashmap usage, consume significantly higher energy than traditional OO anti-patterns.}

%\ruben{Although it is clear, maybe it is good idea to specify what is the answer for each of the two preliminary questions. Maybe splitting this section in two subsections.} \Foutse{Yes, you should explicitly accept or reject your null hypothesis and answer your research questions! } 

\section{Threats to validity}
\label{sec:threats-validity}
This section discusses the threats to validity of our study following common guidelines for empirical studies~\cite{Yin}.

\emph{Construct validity threats} concern the relation between theory and observation. This is mainly due to possible mistakes in the detection of anti-patterns, when applying refactorings. We detected anti-patterns using the widely-adopted technique DECOR~\cite{moha2010decor} and the guidelines proposed by Gottschalk and Android guidelines for developers~\cite{gottschalk2013energy,AndroidPerformance}. However, we cannot guarantee that we detected all possible anti-patterns, or that all those detected are indeed true anti-patterns. Concerning the application of refactorings for the case study, we use the refactoring tool support of Android Studio and Eclipse, to minimize human mistakes. In addition, we verify the correct execution of the proposed scenarios and inspect the ADB Monitor to avoid introducing regression after a refactoring was applied.

Considering energy measurements we used the same phone model used in other papers. Plus our measurement apparatus has a higher or the same number of sampling bits as previous studies and our sampling frequency is one order of magnitude higher than past studies. Overall, we believe our measurements are more precise or at least as precise as similar previous studies. As in most previous studies we cannot exclude the impact of the operating system. What is measured is a mix of Android and application
actions. We mitigate this by running the application multiple times and we process energy and execution traces to take into account only the energy consumption of method calls belonging to the app.

\emph{Threats to internal validity} concern our selection of anti-patterns, tools, and analysis method. In this study we used a particular yet representative subset of anti-patterns as a proxy for design quality. Regarding energy measurements, we computed the energy using well know theory and scenarios were replicated several time to ensure statistical validity. As explained in the {\em construct validity} our measurement apparatus is at least as precise as previous measurement setup.

\emph{Conclusion validity threats} concern the relation between the treatment and the outcome. We paid attention not to violate assumptions of the constructed statistical models. In particular, we used a non-parametric test, Mann-Whitney U Test, Cliff's $d$, that does not make assumptions on the underlying data distribution.

\emph{Reliability validity threats} concern the possibility of replicating this study. The apps and tools used in this study are open-source.

%The data sets used in this paper are publicly available online\footnote{http://swat.polymtl.ca/rmorales/tosem2016Android.zip} for replication purposes.

 It is important to notice that the same model of phone and version of Android operating system should be used to replicate the study. In addition, considering the scenarios defined for each application, they are only valid for the \textit{apk} versions used in this study, which are also available in our replication package. The reason is that the scenarios were collected considering approaches based on absolute coordinates and not on the identifier of components in the graphical user interface (GUI). Therefore, if another model of phone is used or the app was updated and the GUI changed, the scenarios will not be valid.
%We share our data and scripts at:\Foutse{please add the link...}

\emph{Threats to external validity} concern the possibility to generalize our results. Our study focuses on 20 android apps with different sizes and belonging to different domains. Yet, more studies and possibly a larger dataset is desirable. Future replications of this study are necessary to confirm our findings. External validity threats do not only apply to the limited number of apps, but also to the way they have been selected (randomly), their types (only free apps), and provenance (one app store). For this reason this work is susceptible to the App Sampling Problem  \cite{Martin:2015:ASP:2820518.2820535}, which exists when only a subset of apps are studied, resulting in potential sampling bias. Nevertheless, we considered apps from different size and domains, and the anti-patterns studied are the most critical according to developers perception~\cite{palomba2014they,hecht:hal-01178734}.

%\Foutse{you should close on something positive...}

\section{Related work}
\label{sec:relatedWork}

In this section, we discuss related works about automated-refactoring, Android anti-patterns, and the energy consumption of mobile apps.

%analysis and anti-patterns detection in mobile apps and related
%work on energy consumption and refactoring.

\subsection{Mobile anti-patterns}

Linares-V{\'a}squez et al. \cite{ICPC-2014-VasquezKMSPG} leveraged DECOR to detect 18 OO anti-patterns in mobile apps. Through a study of 1343 apps, they have shown that anti-patterns negatively impact the fault-proneness of mobile apps. In addition, they found that some anti-patterns are more related to specific categories of apps.

Verloop \cite{Verloop13} leveraged refactoring tools, such as PMD~\footnote{https://pmd.github.io/} or JDeodorant \cite{FokaefsTC07} to detect code smells in mobile apps, in order to determine if certain code smells have a higher likelihood to appear in the source code of mobile apps. In both works, the authors did not considered Android-specific anti-patterns.

Reimann et al. \cite{ReimannBA14} proposed a catalogue of 30 quality smells specific to the Android platform. These smells were reported to have a negative impact on quality attributes like efficiency, user experience, and security. Reimann et al. also performed detections and corrections of certain code smells using the REFACTORY tool \cite{ReimannSA13}. However, this tool has not been validated on Android apps \cite{hecht:hal-01178734}.

%Also, Yamashita and Moonen \cite{YamashitaM13} have demonstrated
%through a  study with  developers  that the majority of
%them are concerned about code smells. A similar user study have been conducted by Palomba {\em et al.} \cite{palomba14} about bad smells and potential design problems, their nature and severity.
%
%Recenlty, Tufano et al. \cite{ICSE-TufanoPB-15} performed a study to
%investigate how code smells are introduced in
%software, by analyzing change history. The results of their study have revealed that
%most code smells are introduced when there is a creation of files, development of new features, or the enhancement of existing ones.

Other researchers \cite{hecht:hal-01178734} have analyzed the evolution of the quality of mobile apps through the analysis of 3,568 versions of 106 popular Android apps from the Google Play Store. They used an approach, called \textit{Paprika}, to identify three object-oriented and four Android-specific anti-patterns from the binaries of mobile apps.

\subsection{Energy Consumption}

There are several works on the energy consumption of mobile apps \cite{ICSME-2015-AggarwalHS,PolatoBHK15,PangHAH15,ZhangHG14,RasmussenWH14,hindleGreenMiner}.

Some studies proposed software energy consumption frameworks \cite{hindleGreenMiner} and tools~\cite{ICSME-2015-AggarwalHS} to analyze the impact of software evolution on energy consumption.

%There are two ways of estimating the energy consumption of a mobile device, using a hardware setting which is known to be more accurate and software estimation using energy models.

\textit{Green Miner} \cite{hindleGreenMiner} is a dedicated hardware mining software repositories testbed. The \textit{Green Miner} physically measures the energy consumption of Android mobile devices and automates the reporting of measurements back to developers and researchers. A \textit{Green Miner} web service\footnote{https://pizza.cs.ualberta.ca/gm/index.py} enables the distribution and collection of green mining tests and their results. The hardware client unit consists of an \textit{Arduino}, a breadboard with an \textit{INA219} chip, a \textit{Raspberry Pi} running the \textit{Green Miner} client, a USB hub, and a \textit{Galaxy Nexus} phone (running Android OS 4.2.2) which is connected to a high-current 4.1V DC power supply. Voltage and amperage measurement is the task of the \textit{INA219} integrated circuit which samples data at a frequency of $50\, Hz$. Using this web service, users can define tests for Android apps and run these tests to obtain and visualize information related to energy consumption.

Energy models can be provided by a \textit{Software Environment Energy Profile (SEEP)} whose design and development enables the per instruction energy modeling. Unfortunately, it is not common practice for manufacturers to provide \textit{SEEPs}. Because of that, different approaches have been proposed to measure the energy consumption of mobile apps. Pathak et al.~\cite{Abhinav12} proposed \emph{eprof}, a fine-grained energy profiler for Android apps, that can help developers understand and optimize their apps energy efficiency. In \cite{hao_estimating_2013}, authors proposed the software tool \textit{eLens} to estimate the power consumption of Android applications. This tool is able to estimate the power consumption of real applications to within 10\% of ground-truth measurements. One of the most used energy hardware profilers is the \textit{Monsoon Power Monitor} which has been used in several works. By using this energy hardware profiler a qualitative exploration into how different Android \textit{API} usage patterns can influence energy consumption in mobile applications has been studied by Linares-Vasquez et al. \cite{linares-vasquez_mining_2014}.

Other works aimed to understand software energy consumption \cite{PangHAH15}, its usage \cite{SahinPC14}, or the impact of users' choices on it \cite{ZhangHG14,saborido-optimizing-2016}.

Da Silva et al. \cite{Silva2010} analyzed how the inline method refactoring impacts the performance and energy consumption of three embedded software written in Java. The results of their study show that inline methods can increase energy consumption in some instances while decreasing it in others.

Sahin et al. \cite{SahinCGCKPW12} investigated how high-level design decisions affect an application's energy consumption. They discuss how mappings between software design and power consumption profiles can provide software designers and developers with insightful information about their software power consumption behavior.

Pinto et al. \cite{Pinto15} have suggested a refactoring approach to improve the energy consumption of parallel software systems. They manually applied this refactoring approach to 15 open source projects and reported an energy saving of 12\%. %of energy for each one of them.

Researchers \cite{SahinPC14} have investigated the impact of six commonly-used refactorings on 197 apps. The results of their study have shown that refactorings impact energy consumption and that they can either increase or decrease the amount of energy used by an app. The findings of \cite{SahinPC14} also highlighted the need for energy-aware refactoring approaches that can be integrated in IDEs.

Hecht et al. \cite{Hecht:2016:ESP:2897073.2897100} conducted an empirical study focusing on the individual and combined performance impacts of three Android performance anti-patterns on  two open-source Android apps. These authors evaluated the performance of the original and corrected apps on a common user scenario test. They reported that correcting these Android code smells effectively improves the user interface and memory performance.
%There are also several works on refactoring in the context of energy consumption. For example,  researchers  Other works put efforts into suggesting refactorings to enhance energy efficiency \cite{PintoSF15,Gottschalk+2013}.

Recently, researchers \cite{PintoSF15} have examined research results published in top software engineering venues %.  They identified and discussed 12 contributions
%that can be extended into refactoring tools used to
%improve software energy efficiency. The results of this work showed that Mobile apps is the topic having the highest
%number of opportunities (6 out of 11). They also highlighted the
and highlighted the need for more studies that deal with software energy consumption issues. %In \cite{Gottschalk+2013}, the authors proposed an approach, based on code analysis, to reduce energy consumption of applications.
%More recently, researchers \cite{Gottschalk_savingenergy} have suggested an approach to save energy on mobile devices by performing refactoring.
%Their approach involves the definition, detection, and restructuring of energy-inefficient code.

Our work contributes to fill this gap in the literature. % with the following contributions:

%\begin{itemize}
%	\item An empirical study of the impact of anti-patterns on the energy efficiency in mobile apps.  We found that 1) anti-patterns affect energy efficiency, and 2) android anti-patterns have a higher impact on energy efficiency.
%	\item   Based on these results, we propose an automatic refactoring approach that takes into account the energy efficiency of mobile apps. More specifically, we formulate the problem of refactoring as a combinatorial optimization problem with two conflicting objectives to satisfy: anti-patterns correction and energy efficiency.
%	\item We evaluate the effectiveness of our approach at preserving high-level design quality attributes such as understandability, reusability, and extendibility.  We also validate the energy efficiency of the refactoring solutions proposed by our approach on a real mobile phone, and found that we indeed improve energy efficiency for a subset of five randomly selected apps.
%
%\end{itemize}
 %In addition, %ii) Our approach considers energy anti-patterns as well as OO anti-patterns, iii)
 %\ruben{It seems the real phone was only used in the five selected apps for the sanity check. Maybe it is a good idea to rephrase this sentence}. 

%\input{conclusion}

\section{Conclusion and future work}
\label{sec:conclusion}

In this paper we analyze energy consumption of Object-oriented and Energy Anti-patterns in Android. We introduce a novel approach for measuring energy consumption of apps with and without anti-patterns (refactored) and determine the impact of different anti-patterns in a testbed of 59 Android Apps. The results of our empirical evaluation show that in general apps containing anti-patterns consume more energy than those without anti-patterns, and that depending on the type of refactorings applied, is possible to improve or decrease the energy consumption of a mobile app.  The results obtained in these paper are of great value for researchers and practitioners interested in improving the design quality of their apps, and toolsmiths interested in developing automated approaches.

% or


%\appendices
%% you can choose not to have a title for an appendix
%% if you want by leaving the argument blank
%\section{}
%% use section* for acknowledgment
%\ifCLASSOPTIONcompsoc
  %% The Computer Society usually uses the plural form
  %\section*{Acknowledgments}
%\else
  %% regular IEEE prefers the singular form
  %\section*{Acknowledgment}
%\fi

\ifCLASSOPTIONcaptionsoff
  \newpage
\fi



\balance
\bibliographystyle{IEEEtran}
\bibliography{Bibliography,mobile,PURE,mobileRuben}
\end{document}